\documentclass[copyright,creativecommons]{eptcs}
\providecommand{\event}{LTT 2026} 

\usepackage{iftex}

\ifpdf
  \usepackage{underscore}         
  \usepackage[T1]{fontenc}        
\else
  \usepackage{breakurl}           
\fi

\usepackage{preamble}
\usepackage{cleveref}
\usepackage{listings}
\usepackage{xcolor}
\usepackage{amsfonts}
\usepackage{proof}
\lstdefinelanguage{Coq}{
  morekeywords={Definition, Parameter, Fixpoint, CoFixpoint, cofix,Theorem, Proof, Qed, match, with, end, fun, forall, let, in, if, then, else, return, Type, Prop, Set, Inductive, CoInductive, Record, Structure, as, :=, ->, <->, =>},
  sensitive=true,
  morecomment=[l]{(*},
  morecomment=[s]{(*}{*)},
  morestring=[b]",
}

  {\par\noindent\textbf{End of intermezzo.}\par} 

\newcommand{\N}{\mathbb{N}}

\newenvironment{tightlist}%
{\begin{list}{$-$}{%
    \setlength{\topsep}{0in}
    \setlength{\partopsep}{0in}
    \setlength{\itemsep}{0in}
    \setlength{\parsep}{0in}
    \setlength{\leftmargin}{1.5em}
    \setlength{\rightmargin}{0in}
    \setlength{\itemindent}{0in}
}
}%
{\end{list}
}

\lstdefinelanguage{Scheme}{
  morecomment=[l]{;},
  morecomment=[s]{/*}{*/},
  morestring=[b]",
  keywords={%
    define,define-syntax,syntax-rules,define-syntax-rules,abort,%
    let,let*,letrec,%
    lambda,apply,eval,%
    cond,if,else,%
    equal?,eqv?,eq?,%
    quote,unquote,quasiquote,%
    define-record-type,fields,%
    mutable,immutable,%
    callcc,with-exception-handler,%
    match,match*,match**,%
    match-lambda,match-lambda*,match-lambda**,%
    cocase,cocase*,comatch,lambdas,delay%
  }
}
\lstdefinelanguage{Coq}{
  morekeywords={
   Parameter,  Lemma, Theorem, Proof, Qed, forall, fun, Fixpoint, Definition,
    Inductive, CoInductive, match, with, end, as, let, in, return, Type,
    Set, Prop, if, then, else, forall, exists, Constructor, destruct,CoFixpoint,cofix,fix,inl,inr, match, =>
  },
  sensitive=true,
  morecomment=[l]{(*},
  morecomment=[s]{(*}{*)},
  morestring=[b]",
}

\lstdefinelanguage{Coq}{
  morekeywords={Parameter, Definition, Fixpoint, CoFixpoint, cofix,Theorem, Proof, Qed, match, with, end, fun, forall, let, in, if, then, else, return, Type, Prop, Set, Inductive, CoInductive, Record, Structure, as, :=, ->, <->, =>, {|, |}, inl, inr,match},
  sensitive=true,
  morecomment=[l]{(*},
  morecomment=[s]{(*}{*)},
  morestring=[b]",
}

\lstdefinelanguage{OCaml}{
morekeywords={Lazy,lazy,let,rec,if,then,else,match,with,type,of,function,fun,try,raise,exception,in,when,struct,sig,module,open,include,begin,end,object,method,class,val,inherit,new,private,virtual,mutable,as},
  sensitive=true,
  morecomment=[s]{(*}{*)},
  morestring=[b]",
}

\newenvironment{contiguous}[1][\linewidth]
{\par\noindent\minipage{#1}}
{\endminipage\par\noindent\ignorespacesafterend}

\definecolor{codegreen}{rgb}{0,0.6,0}
\definecolor{codegray}{rgb}{0.5,0.5,0.5}
\definecolor{codepurple}{rgb}{0.58,0,0.82}
\definecolor{backcolour}{rgb}{0.95,0.95,0.92}
\lstdefinestyle{mystyle}{
    backgroundcolor=\color{backcolour},
    commentstyle=\color{codegreen},
    keywordstyle=\color{magenta},
    numberstyle=\tiny\color{codegray},
    stringstyle=\color{codepurple},
    basicstyle=\ttfamily\footnotesize,
    breakatwhitespace=false,
    breaklines=true,
    captionpos=b,
    keepspaces=true,
    numbers=left,
    numbersep=5pt,
    showspaces=false,
    showstringspaces=false,
    showtabs=false,
    tabsize=2
}
\title{Proving and Computing:
The Infinite Pigeonhole Principle 
 and 
Countable Choice
}
\author{Zena M. Ariola
\institute{University of Oregon}
\and
Paul Downen
\institute{University of Massachusetts, Lowell}
\and
Hugo Herbelin
\institute{Universit\'e Paris Cit\'e, Inria, CNRS, IRIF}
}
\def\titlerunning{Proving and Computing:
The Infinite Pigeonhole Principle
 and
Countable Choice}
\def\authorrunning{Z. M. Ariola, P. Downen \& H. Herbelin}
\begin{document}
\maketitle

\begin{abstract}
Structural recursion is a common technique used by programmers in modern languages and is taught to introductory computer science students. But what about its dual, \emph{structural corecursion}? Structural corecursion is an elegant technique, supported in languages like Haskell and proof assistants such as Rocq or Agda.  It enables the design of compositional algorithms  by decoupling the generation and consumption of potentially infinite or large data collections. Despite these strengths, structural corecursion is generally considered more advanced than structural recursion and is primarily studied in the context of pure functional programming.

 Our aim is to illustrate  the expressive power of different notions of structural corecursion in the presence of
 classical reasoning. More specifically, we study coiteration and corecursion combined with  the   classical \texttt{callcc} operator, which provides a computational interpretation of
 classical reasoning. This combination enables 
interesting stream-processing algorithms. As an application,   we present a corecursive, control-based proof of
  the \emph{Infinite Pigeonhole Principle} and compare it with the
  continuation-passing proof of  Escard\'o and Oliva in
  Agda. To further demonstrate the power of mixing corecursion and
  control, we give an implementation of the Axiom of Countable Choice.
   In contrast to the usual
  continuation-passing implementations of this axiom,
which rely on general recursion whose  termination is established
  externally, our approach justifies termination by coiteration alone.
\end{abstract}

\section{Introduction}
\label{sec:introduction}

\paragraph{Recursion.}
Recursion on the structure of recursive data types is a common principle for
designing practical programs.  This notion, rooted in induction on
the natural numbers, has been fruitfully applied to programming
for decades. It is now sufficiently well understood that it is routinely taught to beginning computer science students as a general-purpose method for algorithm design \cite{HTDP}.
The advantage of structural recursion is to provide a
template for processing input data of any size by combining the results given for
smaller sub-parts of that input, but with the guarantee that this process will
always finish with a final answer.  For example, the most basic form of
structural recursion is ``primitive recursion'' on numbers---corresponding to
``primitive induction'' and which we refer to as just ``recursion'' for
short---stipulates that recursive calls are only allowed on the immediate
predecessor of the input.  This differs from ``general recursion'', which
imposes no restrictions at all on the recursive calls, and as such does not come
with any guidance for program design or guarantees of termination.  Moreover,
the technique of structural recursion extends far beyond just numbers and can
capture algorithms over any inductively-defined data type, from lists to finite
trees of nearly any shape imaginable.

\paragraph{Corecursion.}
What, then, of the natural dual of structural recursion: \emph{structural
  corecursion}?  Corecursion has been used in several applications of
coalgebras \cite{JacobsRutten97Tutorial,RuttenMethodofCoalgebra}.  But this is
quite different from the way structural recursion is presented, in its own
right, as an independent technique for practical program design. 
Whereas the structure of recursion directly follows the structure of a program's
\emph{inputs}, the structure of corecursion directly follows the structure of
a program's \emph{outputs} \cite{HTDCoPrograms}.  
We aim to show how corecursion provides the right abstraction to compute with
infinite sequences.

\paragraph{Patterns of corecursion.}
Many different formulations of structural recursion 
have their own tradeoffs, creating an impact on issues like ease of
use, expressive power, and computational complexity.  For example, the
recursion scheme corresponding to primitive recursion is sometimes
called a ``paramorphism'' \cite{Meertens92}, with ``catamorphism''
\cite{Hinze13} sometimes used for plain iteration.  Identifying and
studying these various recursion schemes opens a world of laws and
theorems that can be applied by a compiler, such as the catamorphism
fusion laws, which eliminate intermediate structures \cite{Malcolm90}.
Likewise, there are different formulations of structural
corecursions with similar tradeoffs.  Primitive co\-iteration, which
unfolds from a given seed, is called ``anamorphism'' \cite{MFP91};
primitive corecursion, which can stop the computation at any time, is
called ``apomorphism'' \cite{Vene98functionalprogramming}.  These
notions of corecursion are usually studied in the context of
pure functional languages like Haskell
\cite{Unfold99,Graham99,ProofMethodsCorecursive}, where the types of
infinite versus finite objects are conflated.  
In here, instead, we study the
expressive power of various corecursion schemes in the context of the
Rocq proof assistant, and, in particular, in the presence of classical principles.

\paragraph{Infinite Pigeonhole Principle.}
 As an application of corecursion and control, 
we focus on the computational
content of the {\em Infinite Pigeonhole Principle}, which asserts the existence of a constant
subsequence in any sequence of Boolean values $\True$ and $\False$.  The proof is classical,
which means that, computationally, it requires a control operator such
as \texttt{callcc}. The  proof uses corecursion to stepwise
build in parallel the substream of $\True$'s and the substream of $\False$'s: since the
original sequence is infinite, at least one of the substreams produces
arbitrarily many elements.
For comparison, a similar proof expressed in terms of only coiteration does not
produce the same results.
In both cases, we follow the kind of program extraction methodology
studied by Berardi \emph{et al.}~\cite{BarbaneraBerardi93,BerardiCoppoDamianiGiannini00}
to arrive at a program.

\paragraph{Classical Countable Choice.}
The Axiom of Countable Choice asserts that from any total relation on
a countable domain, one can extract a function that selects, for each input,  a witness of
the  relation's totality. The presence of classical
logic increases the strength of the Axiom of Countable Choice. Indeed,
when classical logic is interpreted via a double-negation translation in order to 
retain the constructive character of intuitionistic logic, the Axiom
of Countable Choice expands into the stronger double-negated Axiom of
Countable Choice: $$\forall n.  \neg\neg \exists x^A. R(n,x)
\rightarrow \neg \neg \exists f^{\Nat \rightarrow A} .\forall n . R(n
,f(n)) ~ .$$
 Starting with Spector~\cite{spector1962}, several
implementations of this double-negated form of the Axiom of Countable
Choice have been provided, notably  by Berardi~\emph{et
al.}~\cite{BerardiBezemCoquand1998}. These implementations rely on a form of
general recursion whose termination is postulated. By contrast,
when classical logic is given a direct implementation
by means of control operators is possible, as originally shown by
Griffin~\cite{Griffin90} and later explored, for example, by Berardi \emph{et
al.}~\cite{BarbaneraBerardi93, BarbaneraBerardi95,
  BarbaneraBerardi96}. In this case, general recursion is no longer
needed, and coiteration is enough!

\paragraph{Outline.}
We begin by presenting in Rocq the definition of streams
(Section \ref{streamsRocq}) and continue with different patterns of corecursion
(Section \ref{patternsofcorec}).  We start with coiteration and
then continue with a notion that
allows one to stop the corecursion. We then follow with a more
expressive notion, which, using control, allows one to save a control
point to be resumed later.  
 As an application, we present a 
program that, given a Boolean stream,
returns a constant substream, and 
show that this program corresponds to the
computational content of our proof of the {\em Infinite Pigeonhole
  Principle} (Section \ref{sec:classical-expressiveness}). We then
compare our proof to Escard\'o and Oliva's Agda
proof~\cite{EscardoOliva2011} 
and provide an alternative formulation based on coiteration rather than
corecursion (Section \ref{sec:coiteration-pigeonhole}). We show the strength of
coiteration/corecursion and control by providing a direct proof of the Axiom of Countable Choice
(Section~\ref{coiterationAndControl}).  Before concluding,
we reflect on the different ways to formulate the Infinite Pigeonhole Principle (Section~\ref{Reflections}).
All Rocq code can be found
in the repository  \href{https://github.com/herbelin/infinite-pigeonhole/tree/LTT26}{infinite-pigeonhole}.\\

This work touches on three major contributions of Stefano Berardi:
computing with classical logic, computing with the axiom of countable
choice and extracting programs from proofs. 

\section{Streams in Rocq}
\label{streamsRocq}
%
%
%
%
%
In Rocq, we define a stream in terms of how it is observed or deconstructed:
\begin{contiguous}
\begin{lstlisting}[language= Coq]
CoInductive stream A := { head : A; tail : stream A }.
\end{lstlisting}
\end{contiguous}
The infinite stream of zeros, fitting the above type, becomes:
\begin{contiguous}
\begin{lstlisting}[language=Coq]
CoFixpoint zeroes := {| head := 0; tail := zeroes |}.
\end{lstlisting}
\end{contiguous}
No matter how deep one goes,
\lstinline`(head (tail...(tail (tail zeroes))...))`
will always return~\lstinline`0`.
We can also
generalize \lstinline`zeroes` to any stream that \lstinline`always` returns the
same value~\lstinline`x`:
\begin{contiguous}
\begin{lstlisting}[language=Coq]
CoFixpoint always {X} (x : X) := {| head := x; tail := always x |}.
\end{lstlisting}
\end{contiguous}
or build a stream  by \lstinline`repeat`edly applying some function
\lstinline`f` to a starting value \lstinline`x`:
\begin{contiguous} 
\begin{lstlisting}[language=Coq]
CoFixpoint repeat {X} f (x : X) := {| head := x; tail := repeat f (f x) |}.
\end{lstlisting}
\end{contiguous}
We can also build 
the stream which counts
up starting from some given number \lstinline`n`
(\ie \lstinline`n (n+1) (n+2)...`)
or down
(\ie \lstinline`n (n-1) ... 1 0 0 ...`):
\begin{contiguous}
\begin{lstlisting}[language=Coq]
CoFixpoint countUp n := {| head := n; tail := countUp (n+1) |}.
CoFixpoint countDown n := {| head := n;
                             tail := if Nat.eqb n 0 then zeroes else countDown (n-1)
                          |}.
\end{lstlisting}
\end{contiguous}

\begin{intermezzo}
Since we are interested in extraction, we highlight an issue  with the current Rocq extraction mechanism to call-by-value languages.
We focus on the following simple
 example of a stream of natural numbers, using the evaluation of the Ackermann function on 4 and 4 as a substitute for infinite computation within a language of terminating functions such as Rocq:
\begin{contiguous}
\begin{lstlisting}[language=Coq]
Fixpoint ack n := match n with
                    | 0   => S
                    | S n => fix ackSn m := match m with
                                              | 0   => ack n 1
                                              | S m => ack n (ackSn m)
                                            end
                  end.
CoFixpoint z n := {| head := n; tail := z (ack 4 4) |}.
\end{lstlisting}
\end{contiguous}
A stream, being an infinite object, is not directly observable. However, we can always observe a finite portion of it using an operation that \lstinline`take`s a number  \lstinline`n` of elements:
\begin{contiguous}
\begin{lstlisting}[language=Coq]
Fixpoint take {A} (s : stream A) n :=
  match n with
    | 0    => nil
    | S n' => head s :: take (tail s) n'
  end.
\end{lstlisting}
\end{contiguous}
We would expect that extracting and running \lstinline` take (z 0) 1` would return  \lstinline`0` immediately since we are only interested
in the head of the stream  and not the tail, but instead it computes for a long time.  This is because our  \lstinline`take` function,
even if called with  \lstinline`n` equal to 1,  forces the tail as well. We can understand better
what is going on by extracting the code for the coinductive definition of \lstinline`stream` to OCaml:
\begin{contiguous}
\begin{lstlisting}[language= ml]
type 'a stream = 'a __stream Lazy.t
and 'a __stream =
| Build_stream of 'a * 'a stream
let rec z n = lazy (Build_stream (n, (z (ack (S (S (S (S O)))) (S (S (S (S O))))))))
\end{lstlisting}
\end{contiguous}
The stream is, as expected, a delayed structure. However, according to the call-by-value semantics of OCaml, evaluating 
 \lstinline`(z 0)` forces both components of \lstinline`Build_stream` to be evaluated eagerly; this explains the observed behavior. Indeed, \lstinline` take (z 0) 1` would return immediately if the extraction of \lstinline`z` were instead:
\begin{contiguous}
\begin{lstlisting}[language= ml]
let rec z n =
  lazy (Build_stream (n, lazy (Lazy.force (z (ack (S (S (S (S O)))) (S (S (S (S O)))))))))
\end{lstlisting}
\end{contiguous}

How do we proceed? We are accustomed to delaying the tail in a call-by-value setting and
therefore we ensure laziness ourselves and define \lstinline`stream` as:
\begin{contiguous}
\begin{lstlisting}[language=Coq]
CoInductive stream A := { head : A; tail : unit -> stream A }.
CoFixpoint z n := {| head := n;
                     tail := fun _ => z (ack 4 4)
                  |}.
\end{lstlisting}
\end{contiguous}
\noindent We also redefine our  \lstinline`take` function as
\begin{contiguous}
\begin{lstlisting}[language=Coq]
Fixpoint take A (s : unit -> stream A) n :=
  match n with
    | O    => nil
    | S n' => let s := s tt in head s :: take (tail s) n'
  end.
\end{lstlisting}
\end{contiguous}
And now our   \lstinline`take (fun _ => z 0) 1` does what we expect. 
As another workaround for \lstinline`take`, we could also avoid the delayed argument and use the definition:
\begin{contiguous}
\begin{lstlisting}[language=Coq]
Fixpoint take {A} (s : stream A) n :=
  match n with
    | 0   => nil
    | S 0 => head s :: nil
    | S n'=> head s :: take (tail s tt) n'
  end.
\end{lstlisting}
\end{contiguous}
Note the special case for \lstinline`n=1`; otherwise, the tail of the stream would be  forced, and
we would run into the same problem of evaluating \lstinline`(ack 4 4)` even when we are  only interested in the
\lstinline`head`. 

More generally, one might ask whether the evaluation of the \lstinline`head` shouldn't also be delayed when building a stream.
Indeed, following the idea that a coinductive type is a computation type~\cite{Levy2004}, wouldn't we expect the evaluation of \lstinline`head (tail (z' 4))` to bypass the costly evaluation of the head of \lstinline`z' 4` when only the \lstinline`head` of the tail is requested, as  in the following definition of \lstinline`z'`:
\begin{contiguous}
\begin{lstlisting}[language=Coq]
CoFixpoint z' n := {| head := ack n n;
                      tail := z' (pred n)
                   |}.
\end{lstlisting}
\end{contiguous}
Since our application evaluates the head strictly, we do not delay its evaluation.
\end{intermezzo}

\section{Patterns of corecursion in Rocq}
\label{patternsofcorec}

\paragraph{Coiteration.}
With the use of 
 \lstinline`CoFixpoint` and self-reference, we can 
create a wide variety of infinite streams. However, our goal is to abstract over the different 
ways to recurse.
In all of the previous examples, we recurse  with either
the same value of the actual parameter (like in the  \lstinline`zeroes` stream) or with
an updated value.  This
common pattern of generating a stream is captured by
\lstinline`coiter`ation, which uses an internal \lstinline`state` to
\lstinline`make` the  head on the fly. To generate the tail,
 the \lstinline`state` is \lstinline`updated`  and
 \lstinline`coiteration` continues.
 \begin{contiguous}
\begin{lstlisting}[language=Coq]
Definition coiter {X} {Y} (base:X -> Y) (next:X -> X) : X -> stream Y :=
 cofix f s : stream := {| head := base s;
                          tail := f (next s)
                       |}.
\end{lstlisting}
\end{contiguous}

We can thus rewrite some of the stream definitions seen so far as follows:
\begin{contiguous}
\begin{lstlisting}[language=Coq]
Definition always {X} (x : X) := coiter (fun x=>x)(fun x=>x) x.
Definition countUp n := coiter (fun x=>x)(fun x=>x+1) n.
\end{lstlisting}
    \end{contiguous}

Also, the operation which \lstinline`maps` a function \lstinline`f` over
all the elements of a stream is an instance of \lstinline`coiter`ation:
\begin{contiguous}
\begin{lstlisting}[language=Coq]
Definition maps {X} {Y} (f:X -> Y) xs := coiter (fun xs=>f(head xs))(fun xs=>tail xs) xs.
\end{lstlisting}
\end{contiguous}

\paragraph{Minimal  Corecursion.}
\label{corecursion}
But not every stream can be generated from just \lstinline`coiter`ation.  In fact,
we are tempted to  represent \lstinline`countDown` as
\begin{contiguous}
\begin{lstlisting}[language=Coq]
Definition countDown' n := coiter (fun x=>x)(fun x=>if Nat.eqb x 0 then 0 else x-1) n.
\end{lstlisting}
  \end{contiguous}
 \lstinline`countDown` and \lstinline`countDown'` are \emph{extensionally}  equivalent (only considering what external
observers see), but not \emph{intensionally} (counting other factors used internally
in the implementation).  In particular, \lstinline`countDown'` will keep
checking the value of the internal seed at every step, even after the seed
becomes \lstinline`0` and stops changing further, whereas  \lstinline`countDown` stops the
corecursion and returns the  \lstinline`zeroes` stream.

To more accurately reflect the cost of \lstinline`countDown`, we need a more
general way of generating streams.  This is where the minimal \lstinline`corec`ursor
(also known as apomorphism \cite{Vene98functionalprogramming}) comes into
play: it provides a path for ending the corecursion once the rest of the stream
becomes fully known in advance.
We can allow for an early end to corecursion by having the co\-inductive structure
return a sum (\aka disjoint union \lstinline`stream+X`) of the two possible
options: either return a new seed  of type \lstinline`X` to continue
corecursion, or return a previously-defined stream (of type \lstinline`stream`)
to serve as the remainder.

\begin{contiguous}
\begin{lstlisting}[language=Coq]
Definition corecM {X} {Y} (base:X -> Y) (next:X -> (stream Y) + X) : X -> stream Y := 
            cofix f (s:X) :=  {| head := base s;
                                 tail := match next s with 
                                           |inl s0 => s0
                                           |inr x  => f x 
                                         end
                              |}.
\end{lstlisting}
\end{contiguous}

The definition of \lstinline`countDown` now becomes an instance of minimal corecursion:
\begin{contiguous}
\begin{lstlisting}[language=Coq]
Definition zeroes := coiter (fun x=>x) (fun x=>x) 0.
Definition countDown n := corecM 
                          (fun x=>x)
                          (fun x=>if Nat.eqb x 0 then inl zeroes else inr (x-1))
                          n.
\end{lstlisting}
\end{contiguous}

Another example that shows the need to stop the corecursion is the function that
appends a finite \lstinline`List` in front of an infinite stream:
\begin{contiguous}
\begin{lstlisting}[language=Coq]
Fixpoint append l xs := {| head := match l with 
                                     | []   => head xs
                                     | x::y => x
                                   end;
                           tail := match l with
                                     | []   => tail xs
                                     | _::y => append y xs
                                   end
                        |}.
\end{lstlisting}
\end{contiguous}
This \lstinline`append` function can alternatively be  defined as an instance of minimal
\lstinline`corec`:
\begin{contiguous}
\begin{lstlisting}[language=Coq]
Definition append' {X} (l:list X) xs := corecM 
                       (fun l=>match l with | [] => head xs | x::y => x end)
                       (fun l=>match l with | [] => inl (tail xs) | _::y => inr y end) 
                       l.
\end{lstlisting}
\end{contiguous}
Notice that the bodies of the two functions in
\lstinline`append'` are almost identical to the head and tail branches used in \lstinline`append`.  The difference is 
the co\-inductive step; instead of returning directly when \lstinline`l` is
empty, we must tag the result to indicate the loop has ended.  And
instead of continuing the loop by calling itself when \lstinline`l` is
non-empty, we simply return the updated state \lstinline`y`.

\paragraph{Classical Corecursion.}
\label{classicalcorecursion}
We have seen above that, with a 
sum type, the programmer can control when to end the corecursion.
 But once a corecursive loop is ended, it is done for good.  We would like to do  something more: using 
first-class control, we can provide two
continuations in the \lstinline`tail` step of the corecursive loop; the
first (implicit) continuation lets the \lstinline`tail` step update the state
of the loop and continue corecursing, while the second (explicit) continuation
captures the caller who requested the \lstinline`tail` of the stream.  In other words,
we want the corecursor to have the following type:
\begin{lstlisting}[language=Coq]
corecC : (X -> Y) -> (X -> (stream Y -> B)  -> X) -> X -> stream Y
\end{lstlisting}
The {\tt base} function is the same as before; however, the {\tt next} function
has access to a {\em continuation}. 
How can we define {\tt corecC} in terms of  {\tt corecM}?
Interestingly, in the presence of  classical logic, we  prove the following judgement
$$(\neg S \rightarrow X) \vdash  S + X$$
as follows 
$$\infer[\rightarrow_E]
{\neg S \rightarrow X \vdash S+X}
{\vdash (\neg(S+X) \rightarrow S+X) \rightarrow S+X &
\infer[\rightarrow_I]
{\neg S \rightarrow X \vdash \neg(S+X) \rightarrow S+X}
{\infer[\vee_I]
{\neg(S+X),  \neg S \rightarrow X  \vdash S+X}
{
\infer[\rightarrow_E]
{\neg(S+X), \neg S \rightarrow X \vdash X}
{\neg S \rightarrow X \vdash \neg S \rightarrow X   &
\infer[\rightarrow_I]{\neg(S+X) \vdash
\neg S }{
\infer[\rightarrow_E]
{\neg(S+X) , S \vdash   \bot}
{\neg(S+X)  \vdash  \neg(S+X)  &
\infer[\vee_I]{ S \vdash  S+ X }{S \vdash S}
}
}
}}}}
$$
The  proof relies on the non constructive axiom $$(\neg(S+X) \rightarrow S+X) \rightarrow S+X$$
which is an  instance of \emph{Peirce Law} (named \mbox{\tt PL} in the proof below)
$$(\neg A \rightarrow A) \rightarrow A $$ whose
computational interpretation is the  \lstinline`callcc` control operator, as
described in \cite{DBLP:conf/icalp/AriolaH03}. The above derivation becomes the following program:
{\footnotesize
$$\infer[\rightarrow_E]
{\mbox{\tt next}: \neg S \rightarrow X \vdash \mbox{\tt callcc}(\lambda \mbox{\tt disjret}. inr (\mbox{\tt next}~ (\lambda \mbox{\tt ret}.
\mbox{\tt disjret}(inl~\mbox{\tt ret}))):
S+X}
{\vdash \mbox{\tt callcc}:\mbox{\tt PL}\!\!\!\!\!\!\! &
\infer[\rightarrow_I]
{\mbox{\tt next}: \neg S \rightarrow X \vdash \lambda \mbox{\tt disjret}. inr (\mbox{\tt next}~ (\lambda \mbox{\tt ret}.
\mbox{\tt disjret}(inl~\mbox{\tt ret}))): \neg(S+X) \rightarrow S+X}
{\infer[\vee_I]
{ \mbox{\tt disjret}:\neg(S+X), \mbox{\tt next}:  \neg S \rightarrow X  \vdash
inr (\mbox{\tt next}~ (\lambda \mbox{\tt ret}.
\mbox{\tt disjret}(inl~\mbox{\tt ret})):  S+X}
{
\infer[\rightarrow_E]
{\mbox{\tt disjret}:\neg(S+X), \mbox{\tt next}: \neg S \rightarrow X \vdash \mbox{\tt next}~ (\lambda \mbox{\tt ret}.
\mbox{\tt disjret}(inl~\mbox{\tt ret})): X}
{\mbox{\tt next}: \neg S \rightarrow X \vdash \mbox{\tt next}: \neg S \rightarrow X   &
\infer[\rightarrow_I]{\mbox{\tt disjret}:\neg(S+X) \vdash
\lambda \mbox{\tt ret}.
\mbox{\tt disjret}(inl~\mbox{\tt ret}):\neg S }{
\infer[\rightarrow_E]
{\mbox{\tt disjret}:\neg(S+X) , \mbox{\tt ret}: S \vdash \mbox{\tt disjret}(inl~\mbox{\tt ret}):  \bot}
{\mbox{\tt disjret}:\neg(S+X)  \vdash \mbox{\tt disjret}:
 \neg(S+X)  &
\infer[\vee_I]{ \mbox{\tt ret}: S \vdash inl~\mbox{\tt ret}:  S+ X }{\mbox{\tt ret}:S \vdash \mbox{\tt ret}: S}
}
}
}}}}
$$}

We  arrive at the following definition\footnote{The keyword \mbox{\tt Parameter} is just a type declaration which is deferring the actual definition.} of 
the \emph{classical} \lstinline`corec`:
\begin{contiguous}
\begin{lstlisting}[language=Coq]
Definition cont A := A -> Empty_set.
Parameter callcc : forall {A},  (cont A -> A) -> A.
Parameter throw  : forall {A B}, cont A -> A -> B.
Definition corecC {X} {Y} (base:X -> Y) (next:X -> cont (stream Y) -> X)  : X -> stream Y 
   :=  corecM base (fun s => callcc (fun disjret => 
                                              inr (next s (fun ret => disjret (inl ret))))).
\end{lstlisting}
\end{contiguous}

Unlike a sum type, which definitively ends the loop once and for all,
continuations can be invoked multiple times. This gives the ability to ``pause''
and ``resume'' the loop at will. The difference from the minimal 
\lstinline`corec` above is that in the co\-inductive step, \lstinline`corecC`
captures the caller of the tail projection in the continuation \lstinline`disjret`
and provides it to the \lstinline`next` function along with the current
\lstinline`state`.    Using the classical corecursor, we can
provide an alternative definition of \lstinline`countDown`:
\begin{contiguous}
\begin{lstlisting}[language=Coq]
Definition countDown1 n:= corecC
                          (fun x=>x) 
                          (fun x=>fun k=>if Nat.eqb x 0 
                                           then throw k zeroes
                                           else x-1) 
                          n.

\end{lstlisting}
\end{contiguous}
However, as shown in the next  section, the expressive power of the classical corecursor goes beyond
stopping the corecursion.

\begin{remark}
The classical corecursor cannot be defined using \lstinline`coiter`
combined with \lstinline`callcc` as in:
\begin{contiguous}
\begin{lstlisting}[language=Coq]
Definition corecC {X} {Y} (base:X -> Y) (next:X -> (stream Y -> B) -> X) : X -> stream Y  
 := coiter base (fun s => callcc (fun k => .....)) .
\end{lstlisting}
\end{contiguous}
Note that the continuation \lstinline`k` captures the context that updates
\lstinline`coiter`'s state, \emph{not} the  caller of
\lstinline`next`.  Instead, \lstinline`corecC` provides exactly this extra
information to \lstinline`next`. 
\end{remark}

\section{Infinite Pigeonhole Principle with corecursion} 
\label{sec:classical-expressiveness}


We start   (Section \ref{IPPComputing}) by writing a program that search for the infinite repetitions of
a Boolean in a Boolean stream.  We then extract the program to OCaml\footnote{In the repository we also give the Scheme code} extended with
delimited control.  We then show (Section \ref{InfinitePigeonholePrinciple})  that the program does indeed compute the infinite pigeonhole principle.

\subsection{Computing} 
\label{IPPComputing}
In here, we exhibit the expressive power of classical corecursion over
coiteration: it captures and provides a
  first-class continuation to the tail branch, which is otherwise
  not possible in a pure language.
In particular, the access to the continuation pointing to the tail
caller gives more flexibility than merely stopping the coiteration. 

We are going to write a function, \lstinline`infinite_bool`, which, given a Boolean stream \lstinline`bs`,
returns a stream of indices into \lstinline`bs` such that all indices point to the
same Boolean value.  For example,
 given the following streams:
\begin{contiguous}
\begin{lstlisting}[language=Coq]
Definition always_true := coiter (fun x=>x) (fun x=>x) true.
Definition almost_always_true := {| head := false; tail := {| head := false;
                                                              tail := always_true
                                                           |} 
                                 |}. 
\end{lstlisting}
\end{contiguous}
we would expect \lstinline`infinite_bool always_true` to be bound to the stream of natural numbers (\ie \lstinline`countUp 0`) and  \lstinline`infinite_bool almost_always_true`
to the stream \lstinline`countUp 2`.
However, if we have:
\begin{contiguous}
\begin{lstlisting}[language=Coq]
CoFixpoint alternate := coiter (fun x=>x) (fun x=>negb x) true
\end{lstlisting}
\end{contiguous}
then \lstinline`infinite_bool alternate` could either return the streams of even numbers or
the stream of odd numbers!  This shows that the answer depends on the request,
and this is not surprising since we cannot observe the entire stream.
For example, consider a stream made up of 100 \lstinline`false`'s, followed
by 1 million \lstinline`true`'s, and then infinite \lstinline`false`'s.  If we are
asked for an element that appears 10 times, then we might say \lstinline`false`.
 But then, if we are asked
for 1000 occurrences, we might want to say \lstinline`true` because many more of
them will be found before we see the $101^{st}$ \lstinline`false`.  Yet, if we are
asked for 1 billion occurrences, we have no choice but to say \lstinline`false`;
there simply aren't enough \lstinline`true` occurrences in \lstinline`bs`  to satisfy the
request.

With first-class control, we can 
effectively provide a stream that appears to provide all the indices of the Boolean
occurring infinitely often. 
 Yet, at the end of the
day, only an approximation of it needs to be implemented, because every
terminating program can only inspect a finite number of elements in a stream
before it ends.  The first-class control present in classical corecursion lets
us automatically infer this finite number while the program runs, without the
programmer's explicit knowledge or intervention.
The \lstinline`callcc` operator creates a checkpoint by capturing our observer
in a continuation that we can invoke several times, backtracking to the point
in time when \lstinline`callcc` was called, so we can provide several different
answers to that same observer.  

In this application, we first  \emph{guess} that the head element of the
stream might be the same as the Boolean that  occurs infinitely often. We then create a checkpoint with
\lstinline`callcc`. As long as we keep finding
more repetitions of that head Boolean, then our guess appears correct, and we can
keep providing more indices to repetitions of that Boolean.  However, if we find the
other Boolean in the stream, then our guess might be \emph{wrong}.  In this case, we
can backtrack to the start and change our answer to the other Boolean, continuing
to search into the remainder of the stream.  If we find a repetition of the
first Boolean again, we can backtrack yet again to where we left off originally,
rather than starting over entirely.

\begin{figure}[t]
\centering
\begin{lstlisting}[language=Coq,basicstyle=\ttfamily\footnotesize]
Definition infinite_bool (bs : stream bool) : ipp_stream :=
  let b0 := head bs in                                 (* (head bs) occurs infinitely often *)
  callcc (fun start : cont ipp_stream =>
    coiter
      {ns : {n : nat & stream bool} & cont ipp_stream}  (* the type of the internal state *)
      (fun '(depth; _; _) => depth)                     (* head *)
      (fun '(depth; rest; switch) =>                    (* tail *)
          if bool_dec (head rest) b0
          then (S depth; tail rest; switch)
          else callcc (fun restart => throw switch 
   (* (head bs) does not occur infinitely often *)
        (corecC 
          {n : nat & stream bool}                       (* the type of the internal state *)
          (fun s => projT1 s)                           (* head *)
          (fun '(depth; rest) ret =>                    (* tail *)
              if bool_dec (head rest) b0  
              then throw restart (S depth; tail rest; ret)
              else (S depth; tail rest)
          )
          (S depth; tail rest)))                        (* state of the corecursor  *)
      )
      (0; tail bs; start)).                             (* state of the coiterator *)

\end{lstlisting}
\caption{Search for infinite repetitions of a Boolean in a Boolean stream.}
\label{fig:infinitebool}
\end{figure}

%

This algorithm can be implemented in Rocq, as shown in \cref{fig:infinitebool}.
The \lstinline`stream bool` in input is only observed and we implement it
without delaying the tail. In contrast, to prevent the occurrences of
\lstinline`callcc` from being eagerly evaluated after extraction, it
is important that the  output stream has its tail delayed. For this reason,
we adopt the modified version discussed in
the intermezzo, which we call \lstinline`ipp_stream`. The 
combinators \lstinline`coiter` and \lstinline`corecC` are also
modified accordingly\footnote{The complete Rocq program is
available online at 
\href{https://github.com/herbelin/infinite-pigeonhole/blob/LTT26/IPP_program.v}{IPP\_program.v}.}.
  The top-level \lstinline`infinite_bool` function
takes any Boolean stream \lstinline`bs`.
 The task of \lstinline`infinite_bool` is to return a
stream of natural number indices into \lstinline`bs`, all pointing to the same
Boolean.  To begin, \lstinline`infinite_bool` invokes \lstinline`callcc` to save a
checkpoint for when it was first called, making it possible to completely
\lstinline`start` the stream over from the very beginning, if needed.  Then,
\lstinline`infinite_bool` guesses that there will be enough repetitions of the Boolean
at index 0, named \lstinline`b0`,
 and returns the stream of indices created with coiteration,
where the state is:
\begin{tightlist}
\item the current \lstinline`depth` into the original stream (starting with 0), and 
\item the \lstinline`rest` of the \lstinline`bs` stream to search through, and
\item a continuation to \lstinline`switch` to in case our initial guess is wrong (starting with the
top-level continuation, named \lstinline`start`)
\end{tightlist}
 If the next element is equal to  the \lstinline`head` of the input stream, then we just continue
\lstinline`coiter`ating with an updated state with:
\begin{tightlist}
\item the depth incremented by one, and 
\item  the tail of the  \lstinline`rest` of the stream, and 
\item the same  continuation 
\end{tightlist}
Otherwise, we found a Boolean that
is different from  \lstinline`head bs`, and we have to
\lstinline`switch` our searching mode to look for more occurrences of \lstinline`neg (head bs)`. 
But before  switching, we remember our position in the generation of this stream so that
we can resume it later. We then call the \lstinline`switch` continuation with
a  stream of indices all pointing to  \lstinline`neg (head bs)`.  Interestingly,
this inner stream is an instance of the classical corecursor where the state is:
\begin{tightlist}
\item the current \lstinline`depth` into the original Boolean stream (starting with 1), and
\item the tail of the stream we are searching through
\end{tightlist}
As in the previous stream,  asking for the  \lstinline`head` returns the current \lstinline`depth`.
 For computing the tail, in case the next Boolean is the same as
 \lstinline`neg b0`,  then the state of the corecursion is updated.
 Otherwise, control has  to \lstinline`switch` back to the outer loop, kick-starting it back up.
To do so, it passes an updated state including the new continuation that
\lstinline`return`s to the caller who invoked the tail.

\subsection*{Extracting the program to OCaml with delimited control}

\begin{figure}[t]
\centering
\begin{contiguous}
\begin{lstlisting}[language=OCaml,basicstyle=\ttfamily\footnotesize]
let infinite_bool bs =
  let b0 = head bs in
  callcc (fun start  ->
    coiter 
      (fun pat -> let ExistT (x, _) = pat in let ExistT (depth0, _) = x in depth0)
      (fun pat -> let ExistT (x, switch) = pat in let ExistT (depth0, rest0) = x in
          if bool_dec (head rest0) b0
          then ExistT ((ExistT ((S depth0), (tail rest0))), switch)
          else callcc (fun restart -> throw switch
        (corecC
          projT1
          (fun pat0 ret -> let ExistT (depth1, rest1) = pat0 in
              if bool_dec (head rest1) b0
              then throw restart (ExistT ((ExistT ((S depth1), (tail rest1))), ret))
              else ExistT ((S depth1), (tail rest1)))
          (ExistT ((S depth0), (tail rest0))))))
      (ExistT ((ExistT (O, (tail bs))), start)))
\end{lstlisting}
\end{contiguous}
\caption{Program extracted to OCaml with Delimited control.} 
\label{IPPOCaml}
\end{figure}

In order to run the Rocq program, we extract the code to OCaml, which comes with delimited control\footnote{using the \texttt{delimcc} library},
and define \lstinline`callcc` as \lstinline`shift p (fun k -> k (c (fun x -> abort p (k x))))` where \lstinline`p` refers to the delimiting prompt.
The extracted code \footnote{The complete extracted OCaml program is available 
online at
\href{https://github.com/herbelin/infinite-pigeonhole/blob/LTT26/IPP_program.ml}{IPP\_program.ml}.
We also provide the Scheme code at 
\href{https://github.com/herbelin/infinite-pigeonhole/blob/LTT26/IPP_program.scm}{IPP\_program.scm}.}
 is shown  in \cref{IPPOCaml}.
We test our program with the stream, named  \lstinline`test`,
obtained 
by \lstinline`append`ing
some irregular variations on top of a stream that is \lstinline`always` true:
\lstinline`true false false true false true true true ...`
By using the function  \lstinline`take` of type  \lstinline`stream A` $\rightarrow$ \lstinline`nat` $\rightarrow$  \lstinline`list A`, we can observe
finite prefixes of a stream. For example, 
 if we ask for only three repeated occurrences in the stream \lstinline`test`,
we should observe the first three occurrences of \lstinline`false`:
\begin{lstlisting}[keywordstyle=, commentstyle=, stringstyle=, basicstyle=\ttfamily\footnotesize]
push_prompt p (fun () -> take (infinite_bool test) 3)
- : nat list = [1; 2; 4]
\end{lstlisting}

However, if we ask for five repetitions in that very same stream, there are simply
not enough \lstinline`false` to be found.  Thus, \lstinline`infinite_bool`
will point out five different indices to \lstinline`true`:
\begin{lstlisting}[keywordstyle=, commentstyle=, stringstyle=, basicstyle=\ttfamily\footnotesize]
push_prompt p (fun () -> take (infinite_bool test) 5)
- : nat list = [0; 3; 5; 6; 7]
\end{lstlisting}

Let us now run  both of the above tests on the
\emph{same} stream returned by \lstinline`infinite_bool`:
\begin{lstlisting}[keywordstyle=, commentstyle=, stringstyle=, basicstyle=\ttfamily\footnotesize]
push_prompt p (fun () -> let ix = (infinite_bool test) in [take ix 3; take ix 5]);;
 - : (nat list) list = [[0; 3; 5]; [0; 3; 5; 6; 7]]
\end{lstlisting}
We obtain consistent approximations each time\footnote{Running the above test requires 
defining a new prompt, since the return type is now  \lstinline`(nat list) list`
insead of  \lstinline`nat list`.} . Note that, even though the answer may depend on the observations, a single call to \lstinline`infinite_bool` will always produce \emph{consistent} results throughout its lifetime, regardless of how many times the result is inspected. Even though the first test (requesting only three indices) might initially yield the approximate result  \lstinline`[1; 2; 4]` shown above, it is automatically updated to \lstinline`[0; 3; 5]` to remain consistent with the second, more demanding test (requesting five indices).

\subsection{Proving}
\label{InfinitePigeonholePrinciple}

We successfully wrote the \lstinline`infinite_bool` function  and ran the extracted code, but we do not have any guarantees that
it returns the right sequence of indices; it could as well return the 
\lstinline`zeroes` stream since that satisfies the
property that every index points to the same Boolean. We aim to show that the previous
program constitutes the computational content of the 
 \emph{Infinite Pigeonhole Principle}, which asserts the existence of a constant
subsequence in any sequence of Boolean values $\True$ and $\False$.

\begin{theorem}[Infinite Pigeonhole Principle]
\label{thm:infinite-stream-pigeons}
For every infinite Boolean stream $bs$, there exists  an increasing  stream of indices $s$ such that
each index points to the same truth value in $bs$:
$$\forall bs . \exists s .  \forall n . (s(n) < s (n+1) \wedge  bs(s(n+1)) = bs(s(n))) \enspace . $$

\end{theorem}
Our goal is to embed  this invariant into the definition of the stream.
At the same,  we  want to make sure that the code
extracted from the proof coincides with our original program;
in other words, the program is simply the erasure of the  proof.
It is customary in type
theory to identify propositions with \emph{subsingleton types}, that
is, as the subset of types having at most one element. In this context,
the \emph{propositional truncation}  of a type $A$, \lstinline `trunc A`,  is a
type-theoretic connective which precisely prevents extracting
computational contents from $A$, that is,  \lstinline `trunc A` is a type with at
most one inhabitant expressing that $A$ is inhabited, without being
able to know a precise inhabitant.
Our notion of stream thus becomes:
\begin{lstlisting}[language=Coq]
Inductive trunc A : Prop := Trunc : A -> trunk A.         
Definition ok (bs:trunc (stream bool)) n depth := ......
    (index depth bs' = index n' bs' /\ n' < depth).
CoInductive ipp_stream bs n : Type :=
  { depth : nat;
    rest  : unit -> ipp_stream bs (Trunc (Some depth));
    hyp   : ok bs n depth }.
\end{lstlisting}
Note that the stream is parameterized with respect to the (truncated) Boolean stream in input, named
 \lstinline`bs`,
and the previous index, named  \lstinline`n`.   At  the beginning, the
index is \lstinline`None`, and that is why we have a (truncated) option type.
The \lstinline`depth` refers to the index into the Boolean stream, as in the program, and
the  \lstinline`rest` refers to the tail of the stream.   The  \lstinline`hyp` formalizes the invariant, which,
 apart from looking inside  the box (\ie the $\ldots$ in the definition of \lstinline`ok`), 
 guarantees that all indices point to the same Boolean
  (\lstinline`index depth bs' = index n' bs'`) and that the current index (\ie \lstinline`depth`) is 
  strictly greater than the previous index (\lstinline`n' < depth`).
Our goal then becomes to prove:
\begin{lstlisting}[language=Coq]
infinite_bool bs : ipp_stream (Trunc bs) (Trunc None)
\end{lstlisting}
The Rocq proof is available online at
\href{https://github.com/herbelin/infinite-pigeonhole/blob/LTT26/IPP_proof.v}{IPP\_proof.v}.
We present, in \cref{IPPProgramOCaml}, the OCaml program extracted from the proof, to emphasize that our original program indeed constitutes the computational content of the proof of the Infinite Pigeonhole Principle\footnote{The complete OCaml code is available online
at \href{https://github.com/herbelin/infinite-pigeonhole/blob/LTT26/IPP_proof.ml}{IPP\_proof.ml},
and the Scheme code at \href{https://github.com/herbelin/infinite-pigeonhole/blob/LTT26/IPP_proof.scm}{IPP\_proof.scm}. A proof-irrelevant argument of type \lstinline`trunc nat` remains unerased in the Rocq extraction of \lstinline`coiter` and \lstinline`corecC`. We do not know why, but this explains the presence of the extra \lstinline`\_`s.}.

\begin{figure}[t]
\centering
\begin{lstlisting}[language=Coq,basicstyle=\ttfamily\footnotesize]]
let infinite_bool bs =
  let b0 = head bs in
  callcc (fun start ->
    coiter
      (fun _ pat -> let ExistT (x, _) = pat in let ExistT (depth0, _) = x in depth0)
      (fun _ pat -> let ExistT (x, s) = pat in let ExistT (depth0, rest0) = x in
          if bool_dec (head rest0) b0
          then ExistT ((ExistT ((S depth0), (tail rest0))), s)
          else callcc (fun restart -> throw s
        (corecC
          (fun _ -> projT1)
          (fun _ pat0 ret -> let ExistT (depth1, rest1) = pat0 in
              if bool_dec (head rest1) b0
              then throw restart (ExistT ((ExistT ((S depth1), (tail rest1))), ret))
              else ExistT ((S depth1), (tail rest1)))
          (ExistT ((S depth0), (tail rest0))))))
      (ExistT ((ExistT (O, (tail bs))), start)))
\end{lstlisting}
\caption{OCaml program extracted from the proof}
\label{IPPProgramOCaml}
\end{figure}

\section{Infinite Pigeonhole Principle with coiteration}
\label{sec:coiteration-pigeonhole}

\paragraph{Comparison with the indirect proof of  Escard\'o and Oliva.}
Escard\'o and Oliva~\cite{Escardo2011,EscardoOliva2011}
gave an intuitionistic proof of the double-negated
formulation of the infinite pigeonhole in Agda\footnote{They more precisely prove $\forall f^{\N
  \rightarrow bool}.\neg_R\neg_R\exists b .\exists g^{\N \rightarrow \N}.
\forall n .(g(n) < g(n+1) \land \neg_R\neg_R f(g(n)) = b)$ for $\neg_R
A \defeq A \rightarrow R$, which can be simplified for $R \defeq \bot$ due to
the decidability of equality on Booleans.}:
$$
\forall f^{\N \rightarrow bool}.\neg\neg\exists b .\exists g^{\N
  \rightarrow \N} .\forall n .(g(n) < g(n+1) \land f(g(n)) = b)
$$
Let us fix a value $b_0$. In Escard\'o  and Oliva's proof, it is a fixed value (say $\False$), but
we could also choose, for example,  $b_0 \defeq bs(0)$, as we did in Section~\ref{sec:classical-expressiveness}. Seen in direct style,
the program reasons by
calling the law of excluded middle: either there exists an infinite constant
substream starting with the same value as $b_0$, or, after some
position, the stream becomes constant with the value opposite to
$b_0$. 
Let $A$ be the proposition $\exists n. \forall n'. (n'\geq n \rightarrow bs(n') \not=
b_0)$. Computationally, excluded-middle
sets a backtracking continuation, say $start$, assumes $\neg A$, and
if ever a proof of $A$ is eventually given, returns it to the
backtracking continuation. A substream has to be found in either case,
whether $\neg A$ or $A$ holds.
The case $A$ is direct, taking $g(k)=n+k$. The case $\neg
A$ relies on the Axiom of Countable Choice\footnote{its
double-negated form has a computational interpretation using general recursion in Escardó and Oliva's proof} to
extract a function $f$ such that $\forall n. (f(n)\geq n
\wedge bs(f(n)) = b_0)$. An appropriate $g$ can then be built by
iterating $f$.

Comparing the two proofs leads to the observation that
representing subsequences with streams \textit{vs} functions from $\mathbb{N}$ to $\mathbb{N}$
matters.
Indeed, streams support parameters so that the next elements can depend on
former elements, while the output of a function on some input depends
only on the input itself\footnote{unless going in the direction of
\emph{very dependent functions}~\cite{Hickey1996FormalOI}, the
counterpart of streams in the language of functions}.
For instance, showing that a function $f$ is growing, as requested
in the infinite pigeonhole statement at the top of this section, requires an external property
connecting $f(n)$ and $f(n+1)$, and then proving:
$$\exists f. \forall n. f(n) < f(n+1)$$
Contrastingly,  showing that a stream is
growing can be done by parameterizing the stream with the last visited
value (using  \lstinline`None` at the beginning), directly bundling
the growing condition within the stream:
\begin{lstlisting}[language=Coq]
CoInductive bundled_stream (prev:option nat) : Type :=
  { depth : nat;
    rest  : bundled_stream (Some depth);
    hyp   : if prev is Some prev then prev < depth }.
\end{lstlisting}
 and then proving \lstinline`bundled_stream None`.
The ability to pass parameters to a stream is exploited in the proof
with corecursion in Section~\ref{sec:classical-expressiveness}, where
the continuation given by corecursion is passed along. In Escard\'o and Oliva's proof, a new
subsequence is rebuilt from scratch every time a value different from
$b_0$ is found. Escard\'o and Oliva's proof is asymmetric: it either returns the
subsequence of all values equal to $b_0$, when there are infinitely many of them, or it returns
the trailing subsequence of \emph{consecutive} values equal to $\neg b_0$, starting immediately after
the last occurrence of $b_0$, when there are only finitely many occurrences of $b_0$. In particular, all occurrences of $\neg b_0$
followed by a $b_0$ are eventually skipped (provided we request sufficient many elements to
observe this). A variant of Escard\'o and Oliva's proof using streams
and only coiteration is presented next.

\subsection{A direct proof using coiteration}
\begin{figure}[t]
\centering
\begin{lstlisting}[language=Coq,basicstyle=\ttfamily\footnotesize]
Definition infinite_bool_coiter (bs : stream bool) : ipp_stream :=
  let b0 := head bs in
  callcc (fun start : cont ipp_stream =>
    coiter _
      (fun '(depth; _) => depth)
      (fun '(depth; rest) =>
          if bool_dec (head rest) b0
          then (S depth; tail rest)
          else callcc (fun restart => throw start
        (coiter {n : nat & stream bool}
          (fun st => projT1 st)
          (fun '(depth; rest) =>
              if bool_dec (head rest) b0
              then throw restart (S depth; tail rest )
              else (S depth; tail rest))
          (S depth; tail rest))))
      (0; tail bs)).
\end{lstlisting}
\caption{Infinite Pigeonhole Principle with coiteration.}
\label{IPPcoiter}
\end{figure}

   In \cref{IPPcoiter}, we give a program using coiteration and control that
extracts a constant substream out of a stream $bs$ of $\True$'s or $\False$'s. As in the program given in Figure \ref{fig:infinitebool}, we make use of  \lstinline`ipp_stream` for the output. This
program is inspired by  Escard\'o and Oliva's Agda
proof~\cite{EscardoOliva2011}, using $\Bool$ instead of $\{0,1\}$ and streams in place of functions
from $\mathbb{N}$ to $\Bool$. We can now observe the difference between the two
programs.  If we let   \lstinline`test` be the stream:
$$\False~\True ~\False~ \True ~\True~\True~\True ... $$
if we observe the first four indices, we get the indices
$$ 1~ 3~ 4~ 5$$
with the program using corecursion, but we get
$$3~4~5~6 $$
with coiteration.  The second true value, at index $1$,  is disregarded. 
The proof and  associated program are available online at
\href{https://github.com/herbelin/infinite-pigeonhole/blob/LTT26/IPP_Escardo_proof.v}{IPP\_Escardo\_proof.v}
and  
    \href{https://github.com/herbelin/infinite-pigeonhole/blob/LTT26/IPP_Escardo_program.v}{IPP\_Escardo\_program.v}, respectively.
The extracted OCaml  code is available at 
\href{https://github.com/herbelin/infinite-pigeonhole/blob/LTT26/IPP_Escardo_proof.ml}{IPP\_Escardo\_proof.ml} and the extracted Scheme code is available  at
\href{https://github.com/herbelin/infinite-pigeonhole/blob/LTT26/IPP_Escardo_proof.scm}{IPP\_Escardo\_proof.scm}.

 
%

\section{Countable Choice with coiteration}
\label{coiterationAndControl}
%
The general axiom of choice is formulated as: 
$$ \mathsf{AC} ~:~ \forall x^A .\, \exists y^B .\,  R(x,y) \rightarrow \exists f^{A \rightarrow
  B}.\, \forall x .\,  R(x,f(x))
$$ 

\paragraph{Constructive Choice.} 
The above already needs some clarification: how should $\exists$ and
$\rightarrow$ be interpreted there?
As a logic and programming language, dependent type theory is
generally fine-grained enough to express the differences. For
instance, in type theory, $\exists$ can be possibly interpreted as a
strong, constructive existential quantification, namely a
$\Sigma$-type, equipped with projections $\pi_1 : \Sigma x^A B(x)
\rightarrow A$ and $\pi_2 : \forall z^{\Sigma x^A B(x)} B(\pi_1 z)$ in
which case the axiom of choice, called Martin{-}L\"of's axiom of choice,
or the intensional axiom of choice, is directly derivable, by
distributivity of $\forall$ over $\Sigma$:
$$
\begin{array}{lcl}
  \mathsf{AC^{int}} &:& 
 \forall x^A. \Sigma y^B. R(x,y) \rightarrow \Sigma f^{A \rightarrow
  B}. \forall x . R(x,f(x))\\
    &\defeq& \lambda H.\,(\lambda x. \pi_1(H(x)), \lambda x. \pi_2(H(x)))\\
\end{array}
$$ 


\paragraph{Classical Choice.}
Thanks to propositional truncation, $\exists$ can be interpreted in type theory as the
propositional truncation of $\Sigma$, that is:
$$
\exists x^A. B(x) \defeq  ||\Sigma x^A. B(x) ||
$$
%
In this case, $\exists$ takes its usual propositional meaning
from intuitionistic logic, and the axiom of choice stops being
derivable.
However, relying on Martin{-}L\"of's Axiom of Choice, we can show that 
the statement $\forall x^A . \exists y^B .  R(x,y) \rightarrow \exists f^{A \rightarrow
  B}. \forall x .  R(x,f(x))$ is equivalent to:
$$
{||~||}\mbox{--}\mathsf{Shift} ~:~ \forall x^A . || B(x) || \rightarrow || \forall x^A. B(x) ||
$$
which is reminiscent of the familiar subclassical Double-Negation
Shift principle, usually written $\mathsf{DNS}$, but which we contrastingly
write here $\mathsf{DNS_{Prop}}$:
$$\mathsf{DNS_{Prop}} ~:~ \forall x^A .\neg \neg B(x) \rightarrow
\neg\neg \forall x^A .B(x)$$ when $B$ is a proposition.

In the presence of classical logic, \ie of double negation
elimination or excluded middle for propositions, the propositional
truncation becomes equivalent to double negation:
$$
\neg \neg A \leftrightarrow ||A||
$$
so that, in the presence of classical logic, the axiom of choice is equivalent to:
$$
\mathsf{DNS_{Type}} ~:~ \forall x^A. \neg\neg B(x) \rightarrow \neg\neg \forall x^A .B(x)
$$
which differs from the usual $\mathsf{DNS}$ by the freedom to consider any type
$B(x)$ and not only a proposition.


\paragraph{Classical Countable Choice.}
In the case where $A$ is countable, a computational content for $\mathsf{DNS_{Type}}$
was given by Spector~\cite{spector1962} in the form of
a recursive program that produces increasingly fine approximations of an inhabitant of
$\forall n .B(n)$. From this, one directly obtains a
recursive program for the double-negated form of the countable choice:
$$\mathsf{AC}_\Nat^{\neg\neg} ~:~ \forall n.  \neg\neg \Sigma x^A. R(n,x) \rightarrow
\neg \neg  \Sigma f^{\Nat \rightarrow A} .\forall n . R(n ,f(n))
$$ 
Spector was reasoning in the context of G\"odel's ``Dialectica''
functional interpretation. Later on, Berardi, Bezem, and
Coquand~\cite{BerardiBezemCoquand1998} gave a variant of the
interpretation in the context of Kreisel's modified realizability.
The computational realization of similar principles was also studied
by Berger, Escard\'o and Oliva~\cite{BergerOliva06,EscardoOliva2010a}
and others.
These interpretations differ in how $\mathsf{DNS_{Type}}$
is interpreted (see e.g. Agda file \texttt{K-Shift}
in~\cite{Escardo2011} for their implementation). But, in all cases,
they rely on a general recursion whose termination is 
justified by the Axiom of Dependent Choice or an axiom classically
equivalent to it, such as Bar Induction.

In~\cite{Herbelin12}, the third author describes a formal language
with coinduction \cite{kozen-silva-2017}, control and $\Sigma$-types, $\mathsf{dPA}_\omega$
(avoiding the inconsistencies described in \cite{Herbelin05}), which
provides a control-based computational interpretation of $\mathsf{AC}_\Nat^{\neg\neg}$ internally justified by only coinduction\footnote{See
also Miquey~\cite{Miquey17} for a cross-justification of the
combination of coinduction, control, and $\Sigma$-types in second-order
arithmetic.}.

Indeed, in the statement of $\mathsf{DNS_{Type}}$, $\forall
n .\neg\neg B(n)$ can be interpreted as the stream
$$\{{\tt head} : \neg\neg B(0);  {\tt tail} : \{ {\tt head} : \neg\neg B(1); ... \}\}$$ 
%
and $\neg\neg\forall n.B(n)$ as the
stream $$\neg\neg\{{\tt head} : B(0);  {\tt tail:} \{ {\tt head} : B(1); ...\}\}$$ so that $\mathsf{DNS_{Type}}$ can be rephrased as the distributivity of double negation
over a stream.
Moving to direct style, that is, using \texttt{callcc}, the combination
of coinduction and control explored in \cite{Herbelin12} can be used to
give a coiterative proof of ${||~||}\mbox{--}\mathsf{Shift}$ (and more
generally of $T\mbox{--}\mathsf{Shift}$ for a strong monad $T$)
eventually yielding a proof not only of $\mathsf{AC}_\Nat^{\neg\neg}$
but also of $\mathsf{AC}_\Nat^{T}$, for $T$ any strong monad, and in
particular of $\mathsf{AC}_\Nat^{||~||}$. The Rocq proof is available online at 
\href{https://github.com/herbelin/infinite-pigeonhole/blob/LTT26/ac_coinductive.v}{ac\_coinductive.v}.
  A detailed analysis of this proof  is left for future work.

\section{Different formulations of the Infinite Pigeonhole Principle}
\label{Reflections}
The study revealed that the Infinite Pigeonhole Principle
can be formulated and proved in many different ways. One degree of
freedom when using streams is in bundling the specification in
the definition of the stream, as in the definition of \-
\lstinline`bundled_stream`, which prevents the need for a call to the Axiom
of Countable Choice.
Another degree of freedom
is in inserting double-negation in front of existential quantifiers to
avoid needing classical reasoning, as made by Escardó and
Oliva. Also, in the context of program extraction from Rocq, as we
did, selected parts of the statement can be marked as
non-computational using propositional truncation. As yet another
degree of freedom, the constancy of subsequences, whether they are represented as functions or streams,
can be expressed globally, as in Escardó and Oliva's formalization, in the form:
$$\exists b. \exists g. \forall n. (g(n) < g(n+1) \wedge f(g(n)) = b)$$
or locally,  as in our proof, in the form:
$$\exists g. \forall n. (g(n) < g(n+1) \wedge f(g(n)) = f(g(n+1)))$$

Coming to the difference between the symmetric and asymmetric proofs,
the stronger specification that a subsequence has no holes would be
(whether $g$ is a function or a stream):
$$\exists g. \forall n. (g(n) < g(n+1) \wedge f(g(n)) = b \wedge
\forall n'. ((n'<g(0) \vee g(n)<n'<g(n+1)) \rightarrow f(n') \neq b))$$ while the
stronger statement that a subsequence is constant from some time would
be: $$\exists g. \forall n. (g(n) + 1 = g(n+1) \wedge f(g(n)) = b)$$

Actually,  the infinity of a value $b$ can also be expressed in pure first-order
logic without using subsequences, as in: $$\forall n. \exists n'. (n'
\geq n \wedge f(n') = b)$$ In particular, it seems reasonable to think
that the ``symmetric'' statement $$\exists b. \forall n. \exists
n'. (n' \geq n \wedge f(n') = b \wedge \forall n''. (n\leq n''<n'
\rightarrow f(n'') \neq b))$$ is provable without corecursion, so that
a symmetric proof of Theorem~\ref{thm:infinite-stream-pigeons} would
eventually be possible using countable choice and, hence, only coiteration.

\section{Conclusion}
We presented a direct style proof of the Infinite Pigeonhole Principle
using corecursion and classical reasoning, and began  to compare it
to Escard\'o and Oliva's indirect proof via double-negation
translation.
We also provided a program that interprets the distributivity of
propositional truncation over streams, and thus the Axiom of Countable
Choice, using only control and coiteration.
Clarifying the general
conditions under which control, coiteration, and corecursion can be combined
into a consistent logical system  is left for future work.
A more rigorous study of the different
formulations of the  Infinite Pigeonhole Principle and
 their proofs deserve  to be conducted.

\bibliographystyle{eptcs}
\bibliography{rec}
\end{document}